\documentclass[aps,twocolumn,superscriptaddress,showpacs, nofootinbib]{revtex4}

\usepackage{graphicx}
\usepackage{bm}

\usepackage{natbib}
\usepackage{float}
\usepackage{multirow}
\usepackage{array}
\usepackage{color}
\usepackage{physics}
\usepackage{siunitx}
\usepackage[colorlinks=true, urlcolor=blue, citecolor=black, linkcolor=black, pdfborder={0 0 0}]{hyperref}

\begin{document}

\title{Tunable magnetism of Boron Imidazolate-based Metal-Organic Frameworks}

\author{Jackson Davis}
\affiliation{Department of Physics and Astronomy, Johns Hopkins University, Baltimore, Maryland 21218, USA}
\author{Pilar Beccar-Varela}
\affiliation{Department of Chemistry, Johns Hopkins University, Baltimore, Maryland 21218, USA}
\author{Soumyodip Banerjee}
\affiliation{Department of Chemistry, Johns Hopkins University, Baltimore, Maryland 21218, USA}
\author{Maxime A. Siegler}
\affiliation{Department of Chemistry, Johns Hopkins University, Baltimore, Maryland 21218, USA}
\author{V. Sara Thoi}
\affiliation{Department of Chemistry, Johns Hopkins University, Baltimore, Maryland 21218, USA}
\affiliation{Department of Materials Science and Engineering, Johns Hopkins University, Baltimore, Maryland 21218, USA}
\author{Natalia Drichko}
\email{drichko@jhu.edu}
\affiliation{Department of Physics and Astronomy, Johns Hopkins University, Baltimore, Maryland 21218, USA}

\date{\today}

\begin{abstract}
    Magnetic metal-organic frameworks (MMOFs), where magnetic metal nodes are connected into a crystal structure by organic linkers, have a potential to host exotic magnetic states. We present a study of bulk magnetic properties of four metal-organic frameworks with {the same} boron imidazolate linkers, Cu-BIF, Co-BIF, Ni-BIF, and {newly synthesized} Zn-BIF, displaying a variety of lattice structures and nontrivial magnetic behaviors. While non-magnetic Zn-BIF provides an offset of magnetic response, magnetic susceptibility measurements of the other three magnetic materials demonstrate the presence of weak magnetic interactions in these MOFs, which differ between materials by sign and size. Cu-BIF, where magnetic nodes are connected into octahedral cages, shows simple paramagnetic behavior.  Triangular lattice Co-BIF shows antiferromagnetic interactions on the order of 1~K, and a spin-crossover-like effect in magnetic susceptibility due to thermal depopulation of excited crystal electric field levels. Magnetic properties of Ni-BIF suggest sizable ferromagnetic interactions. Using  DC/AC susceptibility and variable-field DC magnetization, we detect cluster spin-glass behavior in Ni-BIF and discuss possible microscopic origins of this behavior. This work demonstrates the variety of magnetic properties that are possible with a single organic ligand, and establishes the low energy scale of magnetic interactions through superexchange in boron imidazolate frameworks.
\end{abstract}

\maketitle

\section{Introduction}

Frustrated magnetism is one of the hot topics of current condensed matter research. While one of the ultimate goals is to synthesize and characterize spin liquid materials, where highly interacting spins are not ordered down to low temperatures~\cite{Balents2010,Broholm2020}, many exotic magnetic states and new materials are found as a part of this research. One underexplored source of new magnetic materials are magnetic metal-organic frameworks (MMOFs), in which metal nodes are connected into crystals by organic linkers~\cite{Thorarinsdottir2020}. Due to the versatility of organic chemistry and a relatively easy way to substitute metallic nodes without changing the structure or achieving different structures with the same magnetic nodes and linkers, these materials have the potential to be highly tunable magnets. There are a number of theoretical proposals on the realization of exotic magnetism in these materials~\cite{Jacko2015,Yamada2017,yamada2017crystalline,Jiang2021,zhang2017theoretical,kambe2014redox}, as well as a fair amount of MMOFs synthesized~\cite{Thorarinsdottir2020}, but systematic studies of magnetic properties are rare.

\begin{figure}
    \includegraphics[width=\linewidth]{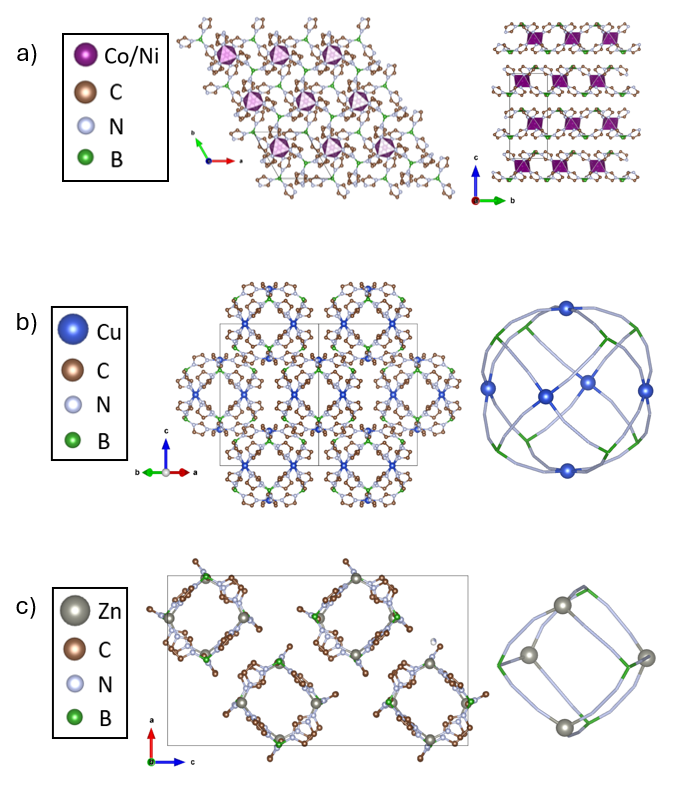}
    \caption{Crystal structure of all four BIFs. {(a) Structure of Co-BIF and Ni-BIF shown along crystallographic c-axis, showing a triangular layer of CoN$_6$ octahedra, and along crystallographic a-axis, showing stacking of layers~\cite{Banerjee2022b}. (b) Packing of Cu-BIF cages and a single Cu-BIF cage schematic with ligands shown as rods~\cite{Banerjee2022a}. (c) Packing of Zn-BIF cages and a single Zn-BIF cage schematic with ligands shown as rods~\cite{Wen2017}. Hydrogen and oxygen omitted from all structures to clearly show metal ions and ligands. All structures visualized in VESTA \cite{Momma2011}.}}
    \label{fig:BIF_structure}
\end{figure}

In this manuscript, we present systematic magnetic studies of four boron imidazolate frameworks (BIFs), {a unique subclass of MOFs that bear resemblances to zeolites and zeolitic imidazolate frameworks~\cite{Zhang2016a, Isaeva2020a},} pictured in Fig. \ref{fig:BIF_structure}. Controlling the synthetic conditions as well as selecting the identity of the metal ion precursor can lead to a variety of solid-state structures. {In this study, we selected 4 BIFs that are supported by the same boron trisimidazolate linker, but with different metal ions.} Co-BIF and Ni-BIF are isostructural and form layered structures of 2D triangular lattices of MN$_6$ octahedra (M = Co (J=1/2), Ni (S=1) ) with $D_{3d}$ symmetry connected by boron imidazolate molecules. Cu-BIF forms a {M$_6$L$_8$ (M = metal ion, L = ligand)} cage structure, with 6 Cu$^{2+}$ ions (S=1/2) in an octahedral arrangement around a central pore. Finally, Zn-BIF forms a different cage structure {(M$_4$L$_4$)}, with 4 Zn$^{2+}$ (S=0) ions in a tetrahedral arrangement around the pore. These structures allow us to explore the effects of metal node substitution and MOF crystal structure on the magnetic properties of MMOFs. 

\section{Methods}

Powders of Co-BIF, Ni-BIF, and Cu-BIF were synthesized according to methods reported in Ref.  \cite{Gerke2023, Banerjee2022a, Banerjee2022b}\footnote{{We name the compounds by their transition metal for clarity; Co-BIF, Ni-BIF, and Cu-BIF were designated CoN$_6$-BIF, NiN$_6$-BIF, and BIF-29, respectively, in ref. \cite{Gerke2023, Banerjee2022a,Banerjee2022b}}}.
{The synthesis of Zn-BIF was adapted from a previous report of a similar Zn boron imidazolate cage~\cite{Wen2017}; experimental details for this newly reported material can be found in the supporting information~\cite{SI}.} Measurements of magnetic properties were performed using a Quantum Design Squid MPMS equipped with the $^3$He cooling option, which allowed measurements down to 0.4 K for selected materials.  DC magnetic susceptibility measurements of powders of Co-BIF, Ni-BIF, Cu-BIF, and Zn-BIF as a function of temperature $\chi$(T) were performed in 2 - 300 K temperature range, with an extension down to 0.4~K for  Co-BIF and Cu-BIF.  All DC susceptibility measurements were performed with a {$\mu_0$}H = 0.1 T applied magnetic field on warming. For the field-cooled measurements the samples were cooled under the same 0.1 T {magnetic} field. {Saturation fields measured for these compounds at minimal temperatures were on the order of 1 T, thus for this comparably small field we calculate magnetic susceptibility $\chi$ as $\chi = M/H$.} Magnetization at 2~K and  above was measured with {magnetic} field sweeps from -7 to 7 T. AC susceptibility measurements of zero-field cooled samples were performed with a 1 Oe AC {magnetic} field amplitude with no applied DC {magnetic} field. For all the measurements, a powder sample of MOF was wrapped in clear plastic wrap and inserted into a plastic straw fixed for the measurements in the MPMS.  Magnetic response of the same wrap in a straw was measured prior to each set of the measurement parameters in order to subtract the background. The total mass of the samples used for the measurements was 5.57 mg Co-BIF, 1.67 mg and 9.05 mg Ni-BIF (9.05 mg data reported here), 3.77 mg Cu-BIF, 22.0 mg Zn-BIF.  The data are normalized to mass/number of moles/number of magnetic ions depending on the measurement.

\section{Results}

\subsection{Zn-BIF}

\begin{figure}
    \includegraphics[width=\linewidth]{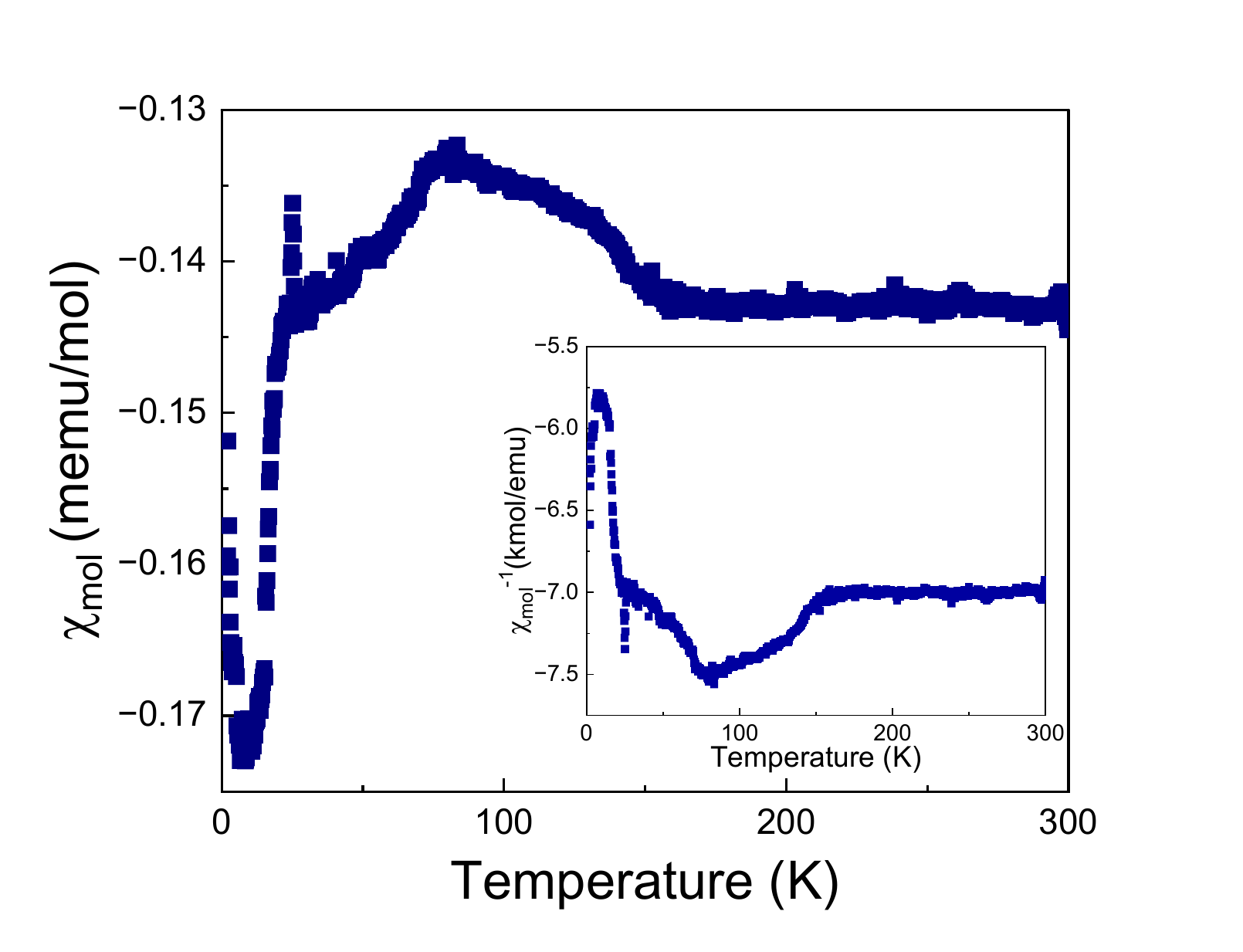}
    \caption{Molar susceptibilty vs. temperature of diamagnetic Zn-BIF. Negative susceptibility that is several orders of magnitude smaller than other BIFs indicates diamagnetism. Deviations from constant susceptibility are attributed to artefacts in background subtraction. Inset: Inverse susceptibility of Zn-BIF vs. temperature.}
    \label{fig:Zn_BIF_MvT}
\end{figure}

With a full{y occupied} 3d electron shell, Zn$^{2+}$ is expected to be {non}magnetic resulting in a diamagnetic response of {Zn-BIF.  The structure of Zn-BIF, shown}  in Fig.~\ref{fig:BIF_structure}{c}, consists of cubic cages, where four Zn ions sit on opposite corners in a tetrahedral arrangement, and each Zn is connected to the others by 2 boron imidazolate linkers. As shown in Fig. \ref{fig:Zn_BIF_MvT}, Zn-BIF displays a negative molar susceptibility that is about 4 orders of magnitude smaller than molar susceptibilities of the other MOFs studied in this work. The deviations from temperature-independent diamagnetic response are a result of the sample magnetization being on the same order of magnitude as the diamagnetic clear plastic wrap background, and thus incurring artefacts as a result of background subtraction.

This measurement suggests that the boron imidazolate linkers themselves are not inherently magnetic, and as a first approximation any magnetic response of other MOFs with the same linker most likely comes from the magnetic metal ions and their interactions mediated by the linkers.

\subsection{Cu-BIF}

\begin{figure}[h!]
    \includegraphics[width=\linewidth]{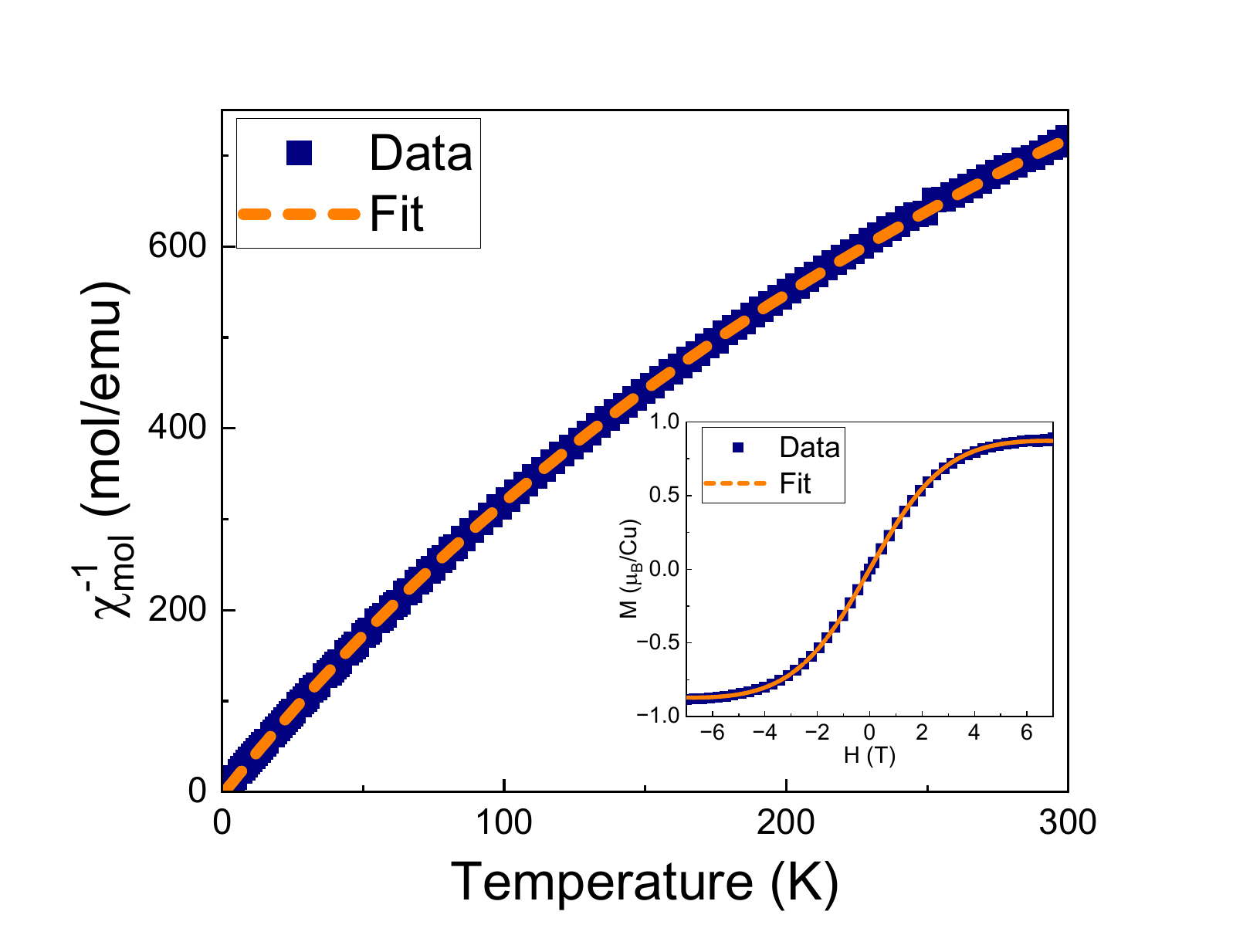}
        \caption{Inverse susceptibility vs. temperature of Cu-BIF, cooled in zero field, fit to a Curie-Weiss paramagnetic model. Inset: Magnetization vs. field of zero-field-cooled Cu-BIF at 2K, fit to Brillouin function for a $S$ = $\frac{1}{2}$ paramagnet.}
    \label{fig:Cu_BIF_magnetic_measurements}
\end{figure}

Cu-BIF exhibits a cage structure which connects each of the 6 magnetic S=1/2 Cu$^{2+}$ ions in an octahedral cage cluster as shown in  Fig. \ref{fig:BIF_structure}b. The Cu$^{2+}$ ions are found in an anisotropic square pyramidal environment, with 4 N from the ligands forming the base of the pyramid and a weakly bound H$_2$O outside the cage forming the tip. Inverse magnetic susceptibility and magnetization of Cu-BIF are shown in Fig. \ref{fig:Cu_BIF_magnetic_measurements}.
Inverse susceptibility is fit to a Curie-Weiss paramagnetic behavior with a weak constant susceptibility $\chi_0$~\cite{Mugiraneza2022}: 

\begin{equation}
    \chi^{-1} = \frac{T-\Theta_{CW}}{\chi_0(T-\Theta_{CW})+C}
\end{equation}

The fit suggests a ferromagnetic Curie-Weiss temperature $\Theta_{CW}$ of {0.9(1)} K, a Curie-Weiss constant $C$ of {0.2587(7)} emu K mol$^{-1}$, and a positive constant offset $\chi_0$ of {0.000526(3)} emu/mol, leading to the slight downward curve in the inverse susceptibility. The constant $C$ = 0.26 emu K mol$^{-1}$ yields an effective moment of 1.44 $\mu_B$, slightly lower than the expected 1.73 $\mu_B$ for Cu$^{2+}$ ions with $S = \frac{1}{2}$ and $g = 2$. No signature of magnetic ordering is present down to T = 0.4 K {(see Fig. 4c of the SI~\cite{SI})}.

Magnetization at T = 2 K is also well fit by a Brillouin function for paramagnetic ions with $S$ = $\frac{1}{2}$ plus a small temperature- and field-independent Van Vleck/diamagnetic susceptibility:

\begin{equation}
    M = gJB_J\left(\frac{gJ\mu_B}{k_BT}H\right)+\chi_0H
\end{equation}

The fit suggests {$g=1.966(1)$}, giving a saturation moment of 0.983 $\mu_B$/Cu, and {$\chi_0=-1.26(1)*10^{-2} \mu_B$/Cu/T}.

\subsection{Co-BIF}

Co-BIF has a layered structure, where  CoN$_6$ octahedra are connected by boron imidazolate (HB(Im)$_3$) molecules into a  triangular lattice forming each layer. (HB(Im)$_3$) linkers connect CoN$_6$  octahedra on the top and bottom of each layer (see  Fig.~\ref{fig:BIF_structure}{a}). 

\begin{figure}
    \centering
    \includegraphics[width=\linewidth]{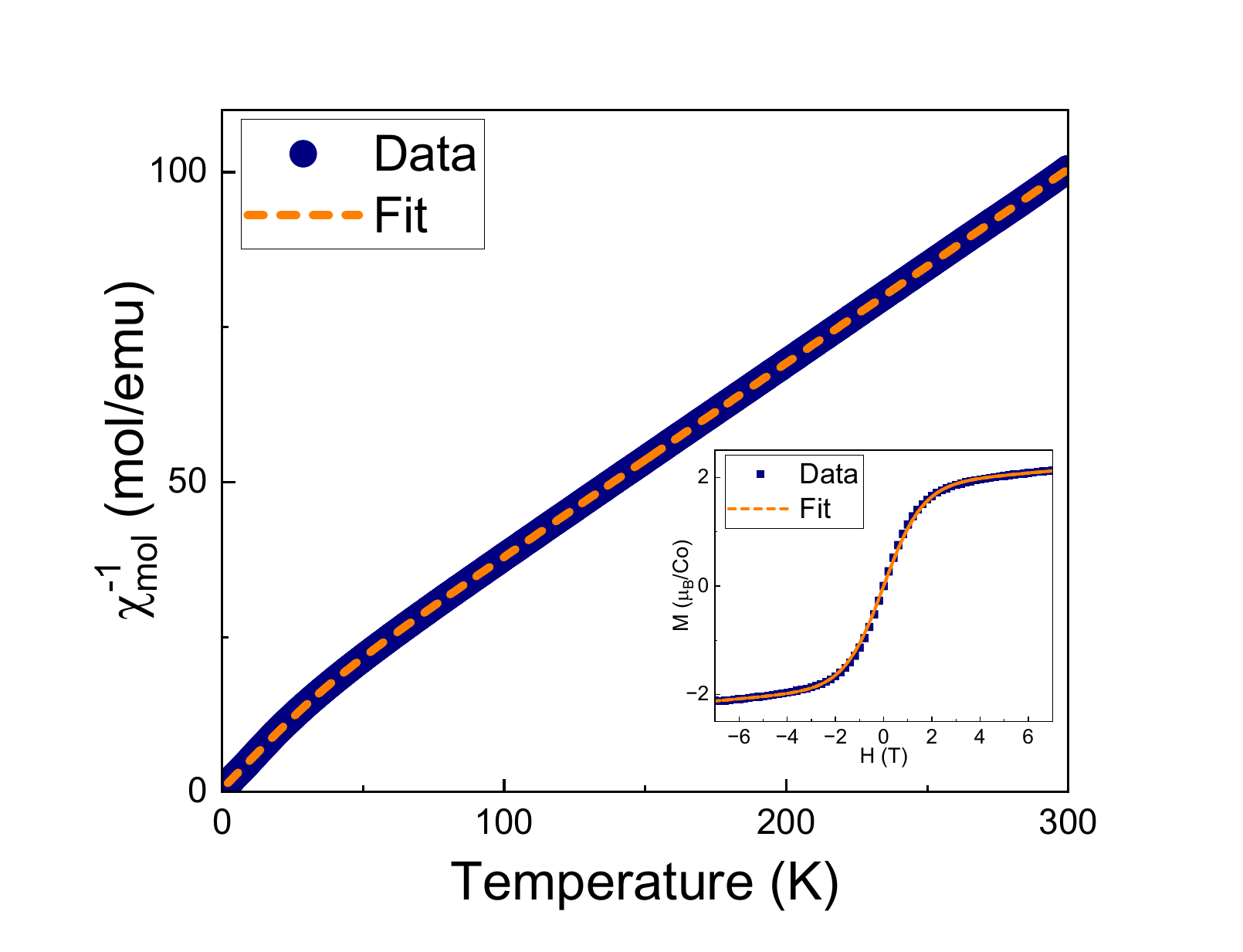}
    \caption{Inverse susceptibility vs. temperature of Co-BIF, fit with a two-level thermal population model, with kink visible around 25 K. Inset: Magnetization vs. field of zero field-cooled Co-BIF at T = 2 K, fit to the  Brillouin function for magnetization of a J = $\frac{1}{2}$ paramagnet.}
    \label{fig:Co_BIF_magnetic_measurements}
\end{figure}

Inverse susceptibility of Co-BIF and magnetization vs. field  are shown in Fig. \ref{fig:Co_BIF_magnetic_measurements}. Inverse susceptibility deviates from Curie-Weiss behavior due to a kink around 25 K, which splits the inverse susceptibility into two linear regions with different slopes. {Following Refs. \cite{Mugiraneza2022}, \cite{Wellm2021}, we fit $\chi^{-1}(T)$ with a model that takes into account the spin-orbit coupling of Co$^{2+}$ and the orbital splitting in the octahedral nitrogen crystal field environment. The crystal electric field relieves the degeneracy of the 3d orbitals, resulting in  a $T_{1g}$ ground triplet which is further split by SOC into a low-spin j = $\frac{1}{2}$ ground state and a high-spin j = $\frac{3}{2}$ excited state.} In a model for magnetic susceptibility, the thermal population of both levels is taken into account. The thermal population model replaces the Curie-Weiss constant {$C = \mu_{eff}^2/8$} with a temperature-dependent factor reflecting the effective moment from thermal population of these states:

\begin{equation}
    \chi^{-1}=8(T-\Theta_{CW})\left(\frac{\mu_{eff, 0}^2 + \mu_{eff, 1}^2e^{\frac{-E_1}{k_BT}}}{1+e^{\frac{-E_1}{k_BT}}}\right)^{-1}
\end{equation}

where $\mu_{eff, 0}$ and $\mu_{eff, 1}$ are the effective moments of the ground and excited states, respectively, and $E_1$ is the energy gap between these states. The fit suggests {$\Theta_{CW} = -1.18(3) K$, effective moments $\mu_{eff, 0} = 4.052(3) \mu_B$ and $\mu_{eff, 1} = 5.734(2) \mu_B$, and a gap $E_1$ = 96.9(3) K.}


Magnetization at T = 2 K is fit to a paramagnetic Brillouin function for J = $\frac{1}{2}$ with a linear contribution from constant susceptibility. The fit suggests {$g=3.78104(5)$}, similar to other Co triangular lattice materials~\cite{Wellm2021}. With $j_{eff} = \frac{1}{2}$, this results in a saturation moment of 1.89 $\mu_B$/Co. The linear contribution is {$\chi_0 = 2.98(3)*10^{-2} \mu_B$/Co/T}, similar in magnitude to Cu-BIF, but positive, indicating a larger Van Vleck paramagnetic contribution.  Following Ref. \cite{Wellm2021}, we additionally fit the data to a delayed Brillouin function, reflecting an exchange field experienced by each Co moment due to magnetization of the sample and exchange interactions with neighboring moments. However, the fit did not significantly improve upon inclusion of this exchange field, reflecting weak exchange interactions between Co magnetic moments.

\subsection{Ni-BIF}

\begin{figure}
    \centering
    \includegraphics[width=\linewidth]{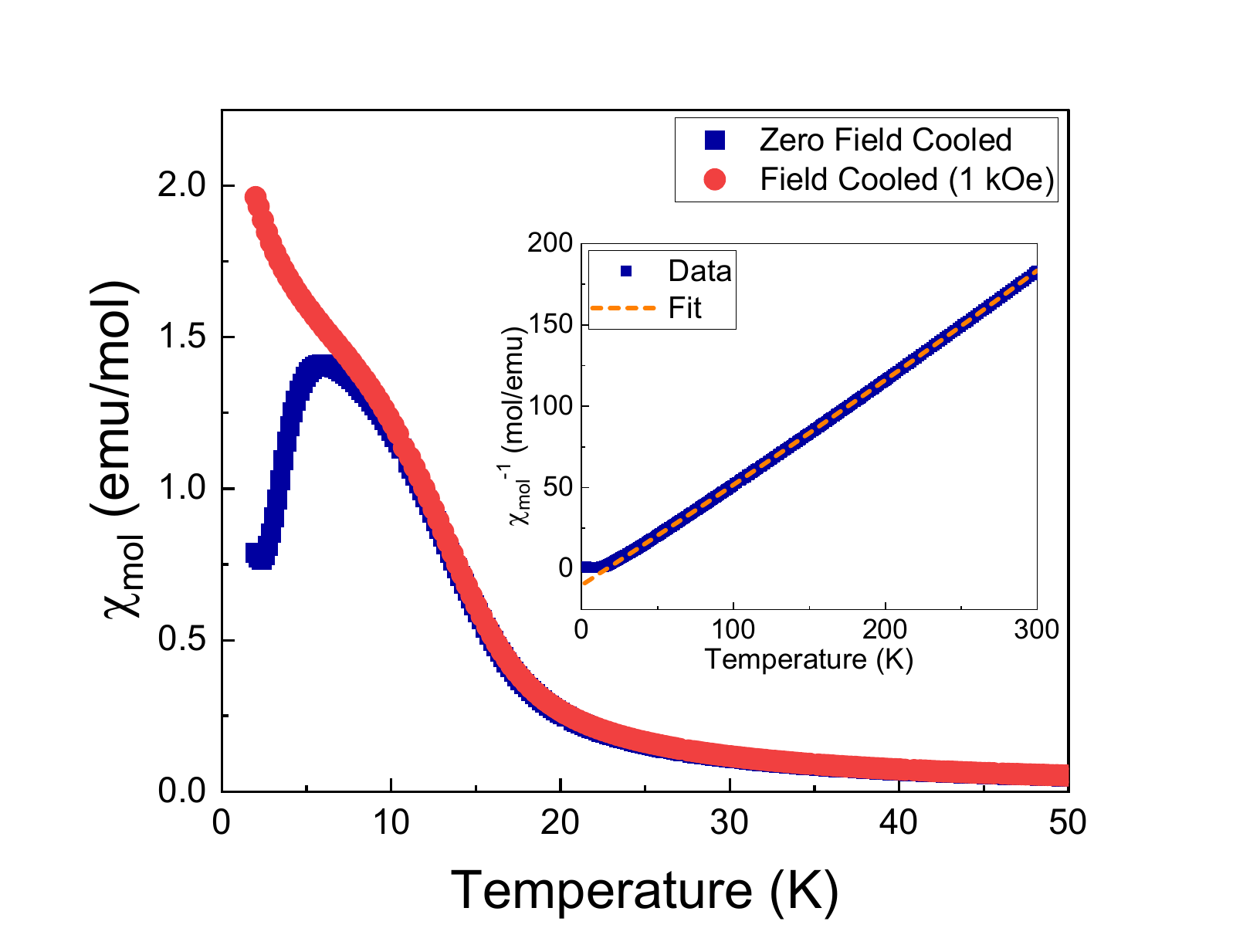}
    \caption{Magnetic susceptibility vs. temperature of Ni-BIF, cooled in both zero field and 1000 Oe applied field. Inset: inverse magnetic susceptibility fit above to Curie-Weiss paramagnetic susceptibility.}
    \label{fig:Ni_BIF_MvT_divergence}
\end{figure}

Ni-BIF, which is isostructural to Co-BIF (see Fig. \ref{fig:BIF_structure}{a}) but contains S = 1 Ni$^{2+}$ ions, displays different magnetic behavior. The inverse susceptibility fit above 100 K suggests  {larger interactions which are ferromagnetic}, with {$\Theta_{CW}=16.5(1)$ K. The fit also yields $C=1.644(2)$}, equating to an effective moment of 3.63 $\mu_B$, {larger than the expected 2.83 $\mu_B$ for S = 1 and g = 2,} and {$\chi_0=-0.000349(6)$} emu/mol. A divergence between zero-field cooled (ZFC) and 0.1 T field-cooled (FC) susceptibility is observed just below 10 K, shown in Fig. \ref{fig:Ni_BIF_MvT_divergence}, with the ZFC susceptibility reaching a maximum at 6 K. 

{Since this result suggests magnetic interactions at least one order of magnitude higher than that in the isostructural Co-BIF, we first performed  tests which confirmed that this result is an intrinsic behavior of Ni-BIF:} (i) To test for possible contamination, magnetic behavior of the chemical precursor, Ni(NO$_3$)$_2 \cdot$ 6H$_2$O, was measured and is shown to be essentially different from the magnetic behavior of Ni-BIF. (see  {Fig. 6d of} the SI~\cite{SI}). (ii) The chemical composition of the powders was confirmed by Raman spectroscopy ~\cite{Davis2023}. (iii) Reproducibility was confirmed for 2 different samples from different synthesized batches. Magnetic {susceptibility} of the two batches is presented in {Fig. 6c of} the SI~\cite{SI}.\\

\begin{figure}
    \centering
    \includegraphics[width=\linewidth]{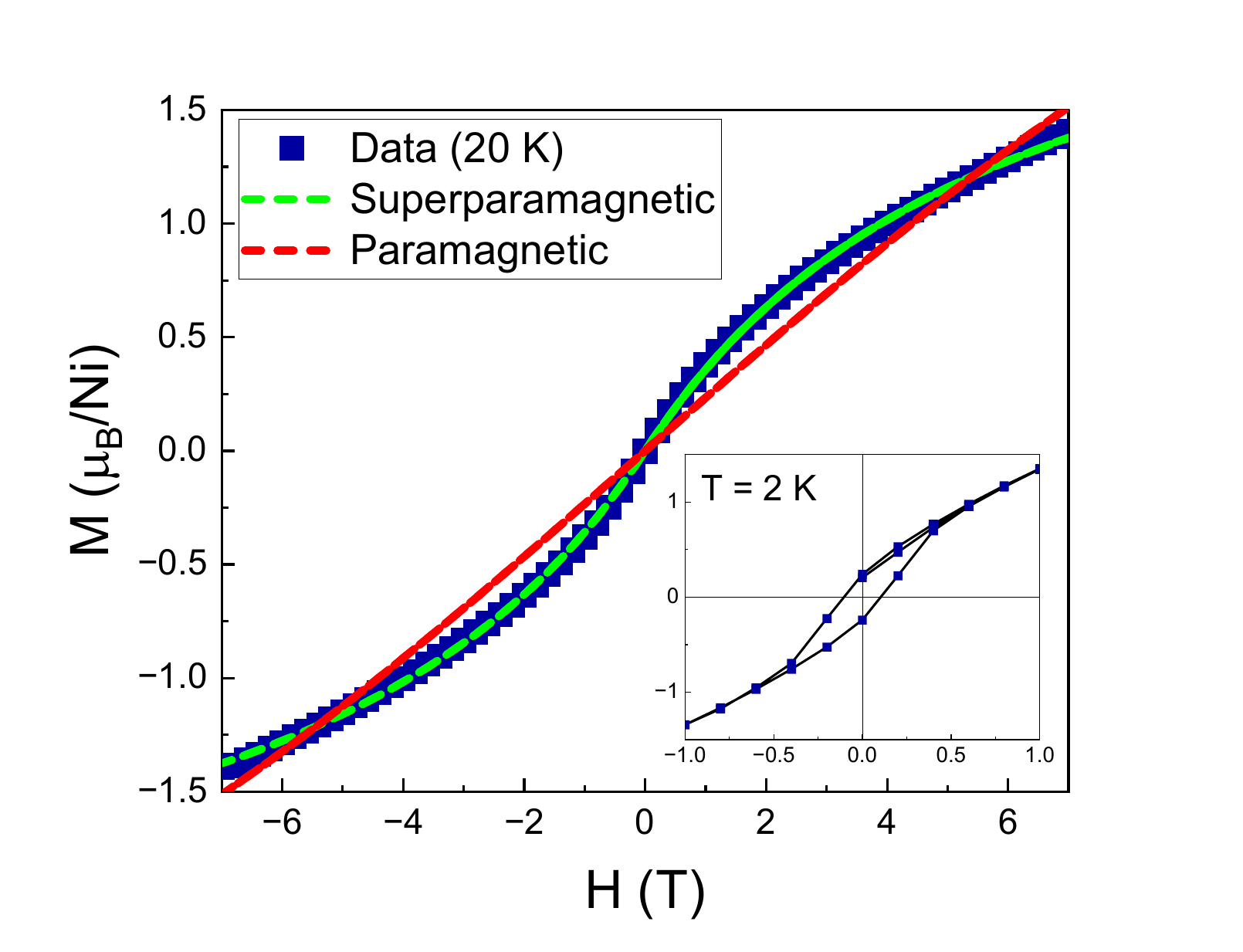}
    \caption{Magnetization vs. field of Ni-BIF at 20 K, fit to a magnetization of a distribution of superparamagnetic nanoparticles, as well as best fit to a Brillouin magnetization function for a J = 1 paramagnet. Inset: Magnetization vs. field at T = 2 K, showing hysteresis.}
    \label{fig:Ni_BIF_MvH_super_20K}
\end{figure}

{Isothermal magnetization also demonstrates a deviation from a paramagnetic behavior in contrast to the other MOFs studied in this work, see Fig. \ref{fig:Ni_BIF_MvH_super_20K}. For magnetization measured at 20 K, above the region of susceptibility divergence, the Brillouin function for S = 1 paramagnetic moments does not fit the measured curve.  Magnetization measured at T = 2 K, below the region of divergence between ZFC and FC susceptibility, recovers paramagnetic behavior on the larger magnetic field scales, but displays a small hysteresis loop with a coercive field of $\sim$0.1 T (see inset in Fig. \ref{fig:Ni_BIF_MvH_super_20K}).}  

The crossover behavior and a difference between FC and ZFC states for magnetic susceptibility, together with magnetization deviating from a Brillouin function at temperatures above the crossover, and a hysteresis present in the low-temperature frozen state are characteristic of cluster spin glasses~\cite{Banerjee2022_CSG,Mydosh_2015}. Here the frozen moments are related to the ordered clusters of spins rather than single spins. Similar to  superparamagnetic heterogeneous nanoparticles, magnetization above the blocking temperature is determined by the presence of a distribution of magnetic moments of different magnitudes~\cite{Ramirez-Meneses2011, Ferrari1997}.



We thus fit the magnetization at T=20~K with a method inspired by Ref. \cite{Ramirez-Meneses2011}. This model assumes a {modified} log-normal distribution of nanoparticle magnetic moments with median moment $\mu_0$ and standard deviation $\sigma$:

\begin{equation}
    f(\mu) = \frac{N}{\sqrt{2\pi}\sigma\mu}\exp\left(-\frac{\ln{\frac{\mu}{\mu_0}}^2}{2\sigma^2}\right)
\end{equation}

{ For superparamagnetic nanoparticles, a magnetization would result in an integration over a continuous range of moments $\mu$, each described by a Langevin function.} 
However, { we obtain the best fit using a discrete sum of Brillouin functions of magnetic moments of small clusters of spins:}

\begin{equation}
    M=\sum_{J=1}^{100} gJB_J\left(\frac{gJ\mu_BH}{k_BT}\right)P(J)
\end{equation}

where $B_J$ is the Brillouin magnetization function for total angular momentum quantum number $J$, and the weight $P(J)$ is given by a discretization of the above log-normal distribution, now using integer $J$ values instead of continuous $\mu$:

\begin{equation}
    P(J) = \int_{J-0.5}^{J+0.5} f(J, J_0, \sigma) dJ
\end{equation}

\begin{figure}
    \centering
    \includegraphics[width=\linewidth]{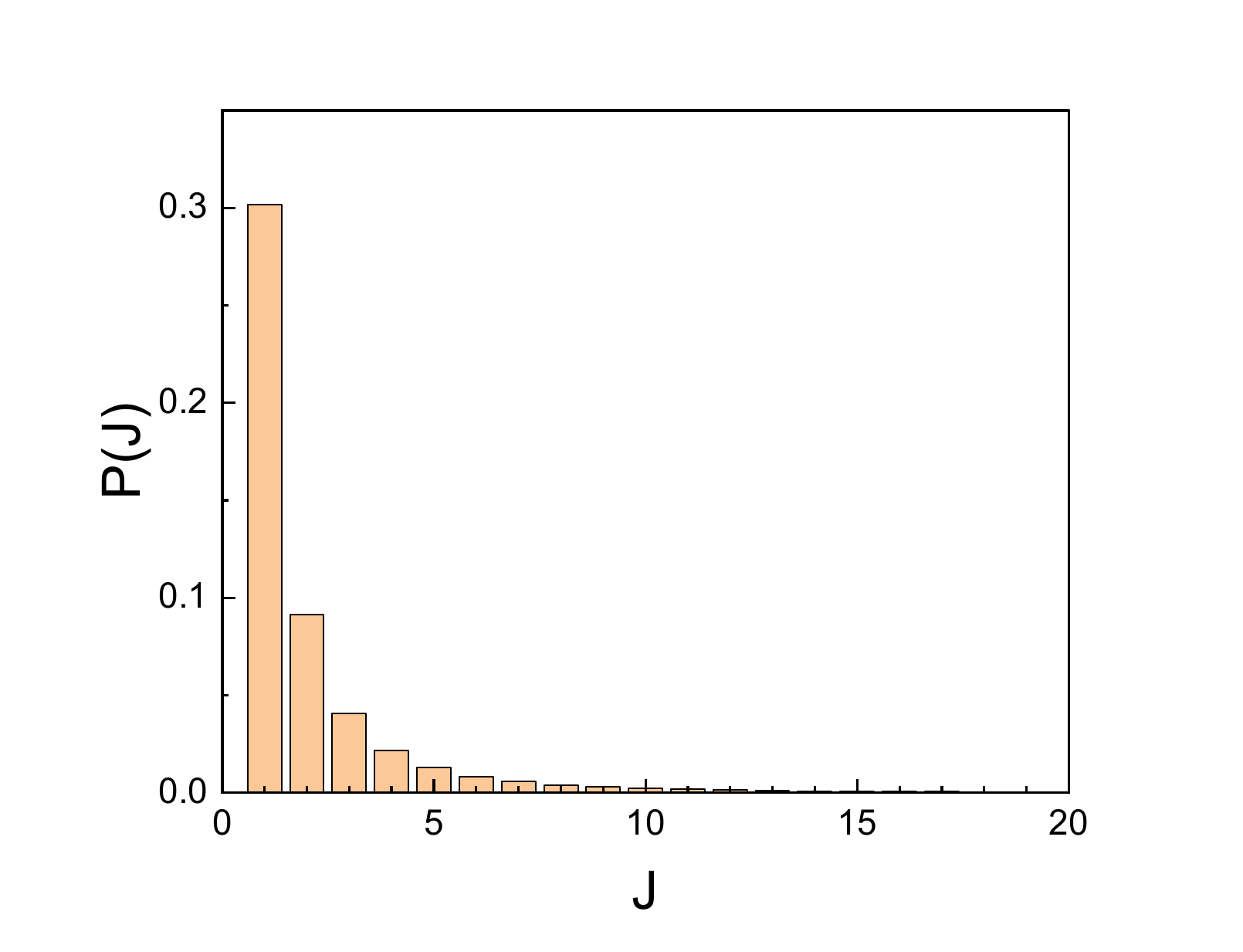}
    \caption{Discrete log-normal distribution of moments in Ni-BIF obtained from fit to T = 20 K magnetization data.}
    \label{fig:Ni_BIF_moment_dist}
\end{figure}

The parameters of the best  fit are {$g = 2.2(8)$}, standard deviation {$\sigma=1.30(2)$} of the quantity $\ln\left(\frac{\mu}{\mu_0}\right)$, and  median moment {$J_0=0.5(2)$} of the continuous log-normal distribution, which equates to a median moment of $J_0=1$ when converted to the discrete distribution shown in Fig.~\ref{fig:Ni_BIF_moment_dist}. The small median moment is an indication that clusters are on the scale of single unit cells, and thus {the possibility of superparamagnetic nanoparticles can be rejected in favor of a cluster spin-glass state. The  distribution of moments that reproduces the observed magnetization is restricted by the modified log-normal distribution used for nanoparticles for fitting purposes; nevertheless, the fit proves that a small percentage of higher moments among a majority of $J=1$ single paramagnetic moments is necessary to describe the data.}

At T = 2 K, below the {crossover}, the magnetization vs. field data are well fit by a single paramagnetic response, with the exception of the hysteresis loop. The fit to a Brillouin function for a $J$ = 1 paramagnet suggests $g_J = 2.400(4)$, with a field-independent linear contribution of $\chi_0 = 8.4(1) * 10^{-2} \mu_B/$Ni/T. The adherence of the data to a single Brillouin function suggests that a single species of moment, the single-spin $J$ = 1 moment, dominates the magnetization response at this temperature.

\begin{figure}
    \centering
    \includegraphics[width=\linewidth]{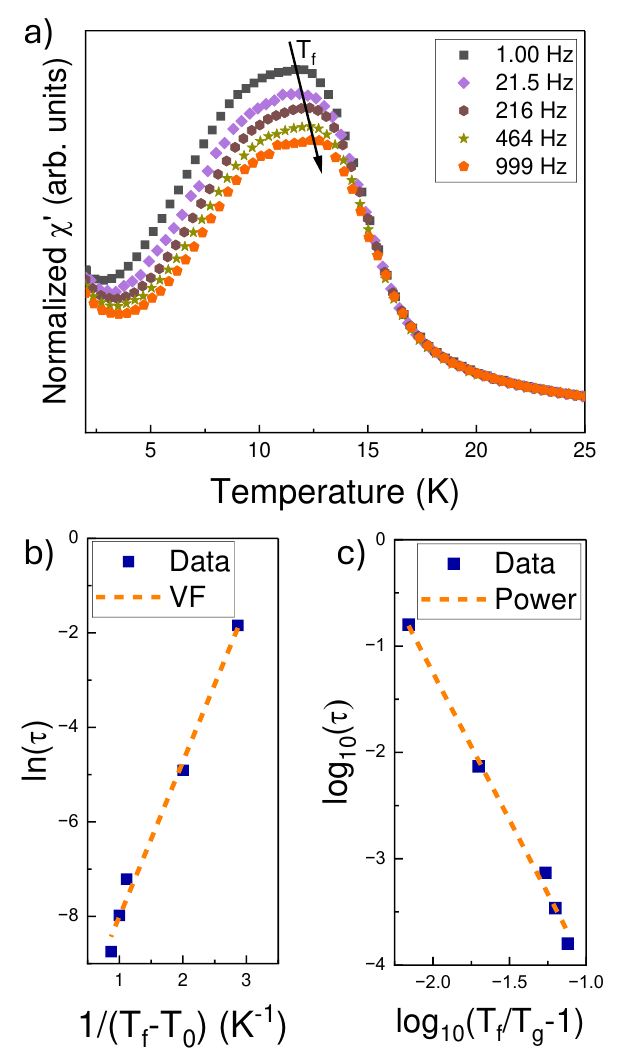}
    \caption{{(a) In-phase component of AC susceptibility of Ni-BIF. $\chi'$ data have been normalized to high-temperature $\chi'$ at 999 Hz; raw data and scaling factors are presented in the SI\cite{SI}. (b) Vogel-Fulcher fit to $T_f$ obtained from $\chi'$. (c) Power law fit to $T_f$ obtained from $\chi'$.}}
    \label{fig:Ni_BIF_AC}
\end{figure}

To further understand the temperature-dependent spin relaxation dynamics of Ni-BIF we performed ZFC AC susceptibility measurements on the same sample. {In-phase susceptibility $\chi'$ is shown in Fig.~\ref{fig:Ni_BIF_AC}a for 5 different frequencies with zero applied DC field. The $\chi'$ data have been normalized to the high-temperature susceptibility at 999~Hz in order to clearly show the evolution of the freezing temperature $T_f$. Raw data are presented in Fig. 7b of the SI~\cite{SI}. $\chi'$ displays a broad, complex peak shape consistent with a broad crossover in DC magnetic susceptibility (Fig.~\ref{fig:Ni_BIF_MvT_divergence}). The peak in $\chi'$ shifts to higher temperatures as frequency is increased. The out-of-phase component $\chi''$, shown in Fig. 7c of the SI~\cite{SI}, highlights the complex peak shape. $T_f$ is approximated from the maximal point on the high-temperature side of the broad peak, as shown in Fig.~\ref{fig:Ni_BIF_AC}a. $T_f$ changes from 11.7 K to 12.5 K between 1 Hz and 999 Hz; thus the Mydosh parameter, $\Delta T_f/(T_f \log_{10}(\Delta \nu))$, or the fractional change in $T_f$ per decade of frequency, is found to be 0.02. This parameter can be used to distinguish between spin-glass and superparamagnetic states, as in conventional spin-glasses it is found to be $\lesssim$ 0.01 while in superparamagnets it is $\gtrsim$ 0.1~\cite{Mydosh_2015, Bag2018}. An intermediate value such as that observed for Ni-BIF is attributed to cluster spin-glass behavior~\cite{Bag2018, Banerjee2022_CSG}.}

{An estimation of characteristic relaxation times allows to distinguish between spin glass and cluster spin glass~\cite{Bag2018, Banerjee2022_CSG}. Relaxation time can be estimated using the Vogel-Fulcher model (see Fig.~\ref{fig:Ni_BIF_AC}b) which adapts the noninteracting Néel-Arrhenius law with a phenomenological parameter $T_0$ accounting for intercluster interactions~\cite{Souletie1985}:}

\begin{equation}
    \tau = \tau_0 \exp(E_a/k_B(T_f-T_0))
\end{equation}

{ where $\tau = 1/2\pi\nu$. The fit suggests a characteristic relaxation time $\ln(\tau_0) = -11(1)$ ($\tau_0 \sim 2.5(2.0) * 10^{-5}$ s), an energy barrier $E_a/k_B$ = 3(2) K, and a VF temperature $T_0$ = 11.2(2) K. Relaxation times of $10^{-7}-10^{-10}$ s are associated with cluster spin-glass, while shorter times of $10^{-12}-10^{-14}$ s are associated with conventional spin-glass systems~\cite{Bag2018}. The timescale derived for Ni-BIF is slower than either of these, but more consistent with cluster spin-glass than spin-glass behavior. }

{An alternative way to estimate relaxation times is to use a power law describing critical slowing down of the relaxation time near the crossover~\cite{Souletie1985}:}

\begin{equation}
    \tau = \tau^* (T_f/T_g - 1)^{-zv'}
\end{equation}

{where $\tau^*$ is a characteristic relaxation time similar to $\tau_0$, $T_g$ is the spin-glass temperature at zero frequency, and the power $zv'$, which is treated as a single value for the fit, is the dynamical critical exponent for the relaxation time near the crossover. The fit in Fig.~\ref{fig:Ni_BIF_AC}c suggests $\log_{10}(\tau^*) = -6.8(7)$ ($\tau^* \sim  4.1(3.8) * 10^{-7}$ s), $T_g$ = 11.62(4) K, and $zv'$ = 2.8(6) K. The suggested ranges of the characteristic relaxation time and the dynamical critical exponent overlap with the ranges of $10^{-7}-10^{-10}$~s and 3-10 typically observed in cluster spin-glass systems~\cite{Banerjee2022_CSG}.}

\section{Discussion}

\begin{table}[ht!]
    \centering
    \begin{tabular}{|c  c  c  c  c|}
         \hline
         BIF & structure & Ion & J & $\Theta_{CW}$ (K) \\
         \hline
         Zn-BIF & cage  &Zn$^{+2}$ & 0 & -- \\
         \hline 
         Cu-BIF & cage  &Cu$^{+2}$ & 1/2 & 0.9\\
         \hline
         Co-BIF & triangular &Co$^{+2}$ & 1/2 & -1.18\\
         \hline
         Ni-BIF &  triangular &Ni$^{+2}$ & 1 & 16.5\\
         \hline
    \end{tabular}
    \caption{Oxidation states, total angular momentum quantum numbers J, and Curie-Weiss Temperatures $\Theta_{CW}$ of BIFs.}
    \label{tab:BIF_table}
\end{table}

The different magnetic properties of the four BIFs suggest the pathways of tuning magnetism of MMOFs by substitution of magnetic metal nodes and synthesis of different crystal structures with the same linker.  Diamagnetism in  Zn-BIF is expected and suggests a ``base line'' to estimate magnetic behavior of magnetic materials of this group. 

Cu-BIF demonstrates robust paramagnetic behavior.  In its structure, each six Cu  atoms are connected  to each other by two boron imadozolate linkers, forming an octahedral cage. While we expect such pathways to provide small magnetic exchange, in agreement with $\Theta_c = 0.9~K$, an absence of magnetic exchange pathways between the cages of 6 Cu atoms provides only a potential possibility for dipole-dipole interactions. Values of $g$, the effective moment, and the saturation moment in Cu-BIF are all slightly suppressed for the expected values for single non-interacting electron spins. 

In the 2D structure of Co-BIF and Ni-BIF, each magnetic node is connected to the other ones by two pathways, ``above'' and ``below'' the layer of octahedra (see Fig. \ref{fig:BIF_structure}). We expect such structure to provide  larger magnetic interactions, while a highly frustrated triangular lattice would prevent magnetic ordering at temperatures close to Curie-Weiss temperature~\cite{Balents2010}. An unexpected result is the ferromagnetic interactions and {cluster spin glass} behavior in Ni-BIF.

Co-BIF is an antiferromagnet with rather low values of magnetic exchange, as evidenced from $\Theta_{CW}$ = -1.18~K. It is interesting to compare this material to  an inorganic triangular lattice of Co octahedra, Na$_2$BaCo(PO$_4$)$_2$.~\cite{Li2020,Zhong2019,Wellm2021}  The Curie-Weiss temperature of $\Theta_{CW}$ = -1.18 K  of Co-BIF is roughly half of that for Na$_2$BaCo(PO$_4$)$_2$, $\Theta_{CW}$ = -2.5 K~\cite{Li2020}, indicating that interactions between spins are reduced by superexchange through boron imidazolate, but still present.  Indeed, while for Na$_2$BaCo(PO$_4$)$_2$ antiferromagnetic exchange $J$ is found to be 1.37 K from the Brillouin function for J = $\frac{1}{2}$  with  exchange field delay, antiferromagnetic exchange in Co-BIF is too small to change purely paramagnetic behavior at 2 K.

Similar to Na$_2$BaCo(PO$_4$)$_2$,  temperature dependence of magnetic susceptibility of Co-BIF is determined by thermal population of low j = $\frac{1}{2}$ and high j = $\frac{3}{2}$ spin states split by {spin-orbit coupling and} crystal field due to trigonal distortion of Co-centered octahedra~\cite{Wellm2021,Mou2024} . The gap between j = $\frac{1}{2}$ and j = $\frac{3}{2}$ in Co-BIF is 96.9 K, which is  roughly 4 times smaller that the gap found in Na$_2$BaCo(PO$_4$)$_2$. This suggests a very small trigonal distortion of the Co octahedra~\cite{Mou2024}, intuitively expected for a large cell and weaker interactions in a molecular crystal. From magnetization measurements done on powders, $g$= 3.78 is close to   4.38 for Na$_2$BaCo(PO$_4$)$_2$. { Using $g=3.78$ obtained from magnetization we calculated expected effective magnetic moments 3.29 and 7.36 $\mu_B$ for the two levels in the population model. While the moment for  $j=\frac{1}{2}$ is close to the observed 4.05 $\mu_B$, the value for  $j=\frac{3}{2}$ is deviating  from the observed 5.73 $\mu_B$. }

Unexpectedly, triangular lattice 2D Ni-BIF, which is isostructural to Co-BIF~\cite{Banerjee2022b}, and shows similar vibrational spectrum ~\cite{Davis2023}, demonstrates very different magnetic behavior. Temperature dependence of magnetic susceptibility shows a  broad peak of a crossover at around T=8~K, with the divergence in susceptibility between ZFC and FC at T=10~K, typical for a spin-glass. {Curie}-Weiss temperature $\Theta_{CW}$= 16.5~K suggests ferromagnetic interactions of much higher magnitude than interactions in the other BIF MOFs.  
Magnetization at T=20~K shows superparamagnetic  {behavior resembling that observed in cluster spin-glass systems or superparamagnetic nanoparticles.}
The small median magnetic moment parameter $J_0 = 1 $, that results from the fit to a {distribution of superparamagnetic nanoparticles}, points to {magnetic clusters on the scale of single unit cells rather than superparamagnetic nanoparticles with diameters of several nm}.
{AC susceptibility measurements provide the estimation of the relaxation times using both Vogel-Fulcher and power law scaling models. Relaxation times of $10^{-5}-10^{-8}$ s are in the range expected for cluster spin glass. Finally, cluster spin glass description of magnetic behavior of Ni-BIF is confirmed by the Mydosh parameter.}

A combination of magnetic frustration, competing antiferro- and ferromagnetic interactions, and disorder is a typical origin of cluster spin glass behavior~\cite{Mydosh_2015}.  While further research is necessary to exactly identify the microscopic origin of relatively large ferromagnetic interactions in triangular lattice Ni-BIF, we can speculate on it. A conversion of weaker antiferromagnetic exchange interactions into one order of magnitude stronger ferromagnetic interactions has been achieved in a magnetic MOF when adding an electron on a linker molecule by doping.~\cite{Perlepe2020}. The extra electron in the radical linker interacting antiferromagnetically with each of the nodes provides the FM exchange between the metal nodes.  Along these lines, we can suggest a concentration of disordered radical spins on some of the linkers. Since each linker binds to three metal nodes in this structure, a single radical could create a cluster of ferromagnetically ordered spins that sits in a lattice of much weaker AFM interactions.





A Raman scattering study of a MOF with a ligand containing an aromatic ring showed that the molecular vibrational modes of the bonds associated with this ring changed significantly upon addition of a radical electron spin~\cite{Liu2019b}. In our previous Raman scattering study of BIFs, we observed no significant change in these high-energy vibrational modes of the imidazole ring~\cite{Davis2023}, indicating that these radicals may be localized to the boron in the linker rather than the imidazolate rings. 


\section{Conclusion}

In this work we demonstrate, using an example of metal-organic frameworks with imidazole-based linkers, robust magnetism of MOFs with magnetic nodes and tunability of magnetic properties. No magnetic response was observed for Zn-BIF with non-magnetic node, while a cage structure of Cu-BIF shows paramagnetism with parameters expected for Cu$^{2+}$ S=1/2 and very weak FM interactions. The triangular lattice structures of Co-BIF and Ni-BIF demonstrate magnetic properties of interest. Co-BIF shows weak AF interactions and single-ion magnetism similar to inorganic materials. Ni-BIF shows sizable FM interactions and {cluster spin-glass behavior which we attribute to} disordered radical spins on the ligand. These results highlight the possibility of leveraging the inherent chemical tunability of MOFs to realize a variety of magnetic systems, especially when using different ligands that mediate stronger superexchange interactions.

\section{Acknowledgements}
The authors are grateful to C. Broholm and K. Thirunavukkuarasu for discussions. Acknowledgement is made to the donors of the American Chemical Society Petroleum Research Fund for partial support  of this research. J. D. and N. D. acknowledge the support of  NSF Award No. DMR-2004074. {P. B.-V., S. B., and V. S. T. acknowledge support by the U.S. Department of Energy (DOE), Office of Science, Office of Basic Energy Sciences, Catalysis Science program, under Award DE-SC0021955. P.B.-V. also thanks the Dean’s ASPIRE grant from the Office of Undergraduate Research, Scholarly and Creative Activity at Johns Hopkins University.}

\bibliography{BIF_Magnetism}

\end{document}


\title{Supplementary Information for Tunable magnetism of Boron Imidazolate-based Metal-Organic Frameworks}
\author{Jackson Davis}
\affiliation{Department of Physics and Astronomy, Johns Hopkins University, Baltimore, Maryland 21218, USA}
\author{Pilar Beccar-Varela}
\affiliation{Department of Chemistry, Johns Hopkins University, Baltimore, Maryland 21218, USA}
\author{Soumyodip Banerjee}
\affiliation{Department of Chemistry, Johns Hopkins University, Baltimore, Maryland 21218, USA}
\author{Maxime A. Siegler}
\affiliation{Department of Chemistry, Johns Hopkins University, Baltimore, Maryland 21218, USA}
\author{V. Sara Thoi}
\affiliation{Department of Chemistry, Johns Hopkins University, Baltimore, Maryland 21218, USA}
\affiliation{Department of Materials Science and Engineering, Johns Hopkins University, Baltimore, Maryland 21218, USA}
\author{Natalia Drichko}
\email{drichko@jhu.edu}
\affiliation{Department of Physics and Astronomy, Johns Hopkins University, Baltimore, Maryland 21218, USA}
\date{May 2024}

\maketitle

In this Supplementary Information we present additional details about {MOF synthesis}, the crystal structure of the studied MOFs as well as  extensive information related to the measurements of their magnetic response. 

{\section{Synthesis and Characterization of Zn-BIF}}
{\subsection{Synthesis of Zn-BIF}}
{All studied MOFs in this report are composed of metal ions in their +2 oxidation state and the same monoanionic boron trisimidazolate linkers, hereby designated as BH(Im)$_3$, where Im = imidazolate. Zn-BIF is a new material and was synthesized as follows: In a 20 mL scintillation vial, 80 mg of Zn(NO$_3$)$_2$ and 50 mg of the sodium form of the boron trisimidazolate ligand, (NaBH(Im)$_3$), was added to a solution containing 3 mL of dimethylacetamide and 2 mL of methanol. The vial was sealed and placed in an oven set at 85 °C for 5 days. The resulting white solid was washed in methanol, followed by acetone, and allowed to dry in air. A single colorless crystal of Zn-BIF was selected for single crystal X-ray crystallography.}

{\subsection{Single Crystal X-ray Crystallography}}

{All reflection intensities were measured at 110(2) K using a SuperNova diffractometer (equipped with Atlas detector) with Cu K$\alpha$ radiation ($\lambda$ = 1.54178 \r{A}) under the program CrysAlisPro (Version CrysAlisPro 1.171.39.29c, Rigaku OD, 2017). The same program was used to refine the cell dimensions and for data reduction. The structure was solved with the program SHELXS-2018/3 (Sheldrick, 2018) and was refined on F2 with SHELXL-2018/3 (Sheldrick, 2018). Analytical numeric absorption correction using a multifaceted crystal model using CrysAlisPro. The temperature of the data collection was controlled using the system Cryojet (manufactured by Oxford Instruments).  The H atoms were placed at calculated positions (unless otherwise specified) using the instructions AFIX 43 with isotropic displacement parameters having values 1.2 Ueq of the attached C atoms.  The H atoms attached to B1X (X = A-D) and O1W were found from difference Fourier maps, and their coordinates were refined pseudofreely using the DFIX instructions in order to keep the B-H, O-H and H-H distances within some acceptable ranges.}

{The structure of Zn-BIF is partly disordered.  The anions coordinated to Zn2 and Zn4 corresponds to a mixture of nitrite (major component) and nitrate (minor component) anion, while there is one disordered nitrate anion coordinated to Zn3.   There is one ordered water molecule coordinated to Zn1.  The fourth anion is found inside the cage-like structure, and is severely disordered over at least three orientations.  All occupancy factors for both major / minor components of the disorder can be retrieved from the final .cif file. The .cif file can also be found on the Cambridge Structural Database (CCDC Number 2374960).}

{\subsection{Powder X-Ray Diffraction Pattern of Zn-BIF}}

\begin{figure}[h]
    \centering
    \includegraphics[width=0.5\linewidth]{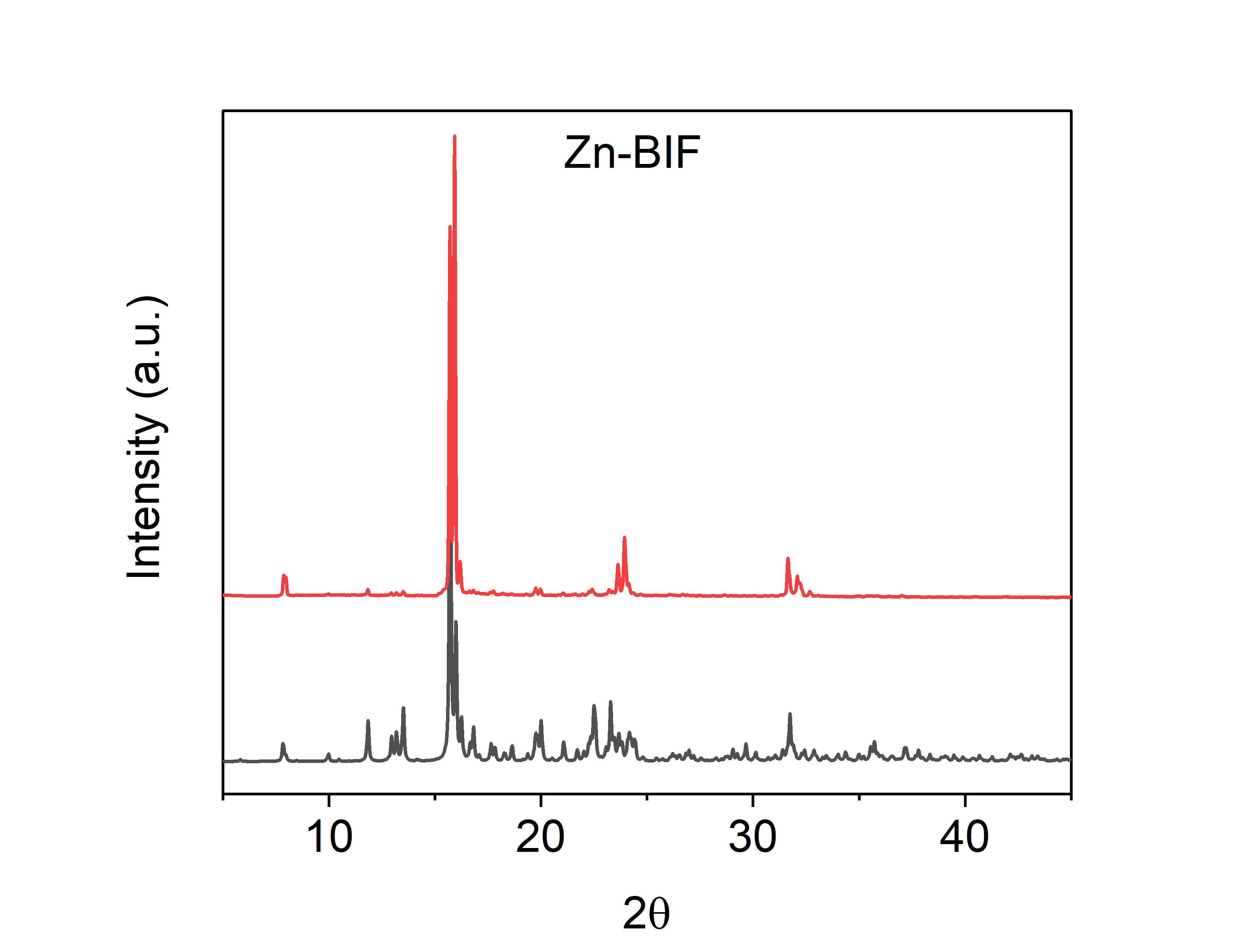}
    \caption{{Powder X-ray Diffraction pattern of Zn-BIF as synthesized (red) and simulated from the single crystal solid-state structure (black).}}
    \label{fig:Zn_BIF_PXRD}
\end{figure}

\section{Structure of the studied MOFs}

\begin{figure}[h]
    \includegraphics[width=\linewidth]{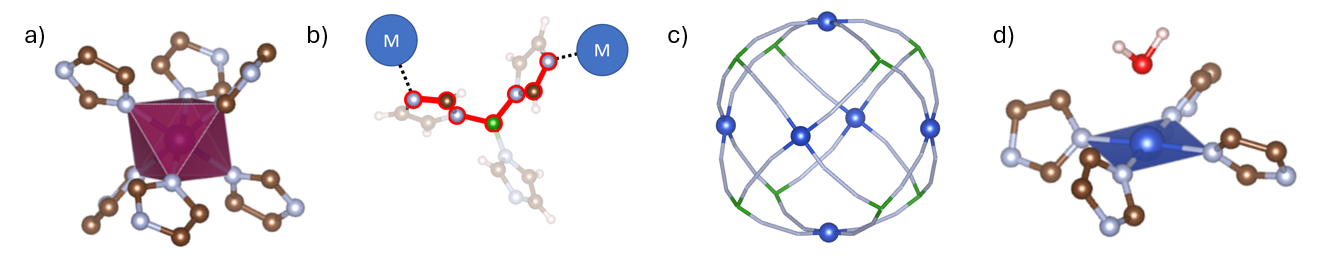}
    \caption{Octahedral environment of Co/Ni in Co/Ni-BIFs (a). Schematic of superexchange path between metal ions through boron imidazolate (b). Simplified schematic of Cu-BIF cage structure showing octahedral arrangement of Cu in each cage. Imidazole rings are represented by 2 N to show connections between Cu (c).  Anisotropic square pyramidal environment of Cu in Cu-BIF (d). Blue = Cu, purple = Co/Ni, light blue = N, brown = C, red = O, pink = H.}
    \label{fig:BIF_SI_structure}
\end{figure}

MOF crystal structures allow in some cases to produce environments of magnetic ions very different from the standard octahedral and tetrahedral environment of magnetic metal ions observed in inorganic magnetic materials. Magnetic BIF-based MOFs studied in this work provide examples of both. The 2D triangular lattice structure of Co-BIF and Ni-BIF presents an example of the standard octahedral environment of  Co and Ni ions, connected by BIF linkers into a triangular lattice. Fig. \ref{fig:BIF_SI_structure} (a) shows this conventional octahedral environment of Co$^{2+}$ and Ni$^{2+}$ in Co- and Ni-BIF formed by 6 imidazole rings. Interestingly, as we show in the main manuscript, the triangular distortion of the octahedral environment estimated for Co-BIF is smaller than that observed in inorganic materials. 


Fig. \ref{fig:BIF_SI_structure} (b) shows the shortest superexchange path between magnetic metal centers through a single boron imidazolate ligand. In all BIFs, total magnetic superexchange between two neighboring metal ions consists of two such paths from two boron imidazolate molecules. 


Cu-BIF provides a very anisotropic environment for Cu, which to the best of our knowledge cannot be achieved in inorganic materials. In Fig. \ref{fig:BIF_SI_structure} (d) we demonstrate this highly anisotropic square pyramidal environment of Cu$^{2+}$ in Cu-BIF formed by 4 imidazole rings and 1 water molecule at the 'tip' of the pyramid. This demonstrates the ability of MOFs to create new environments for magnetic atoms, and related orbital structure can be utilized to tune magnetic properties. Calculations which are beyond the scope of this manuscript are necessary to explore the orbital structure of Cu in such environment. 

In contrast to triangular 2D BIF-MOFs, the structure of Cu-BIF does not provide a continuous network of magnetic atoms connected in a crystal structure. Instead, 6 atoms of Cu are connected into a octahedral-shaped cage, as demonstrated in  Fig. \ref{fig:BIF_SI_structure} (c). This figure shows a simplified version of the Cu-BIF cage structure with carbon removed from imidazole rings to show a minimal schematic of cage structure and connections between Cu. Therefore, magnetic interactions which determine magnetic properties should be the ones between the clusters, not connected by the linkers, and are expected to be very weak. 

\newpage

\section{Magnetic susceptibility}

In this section we report supporting information on magnetic susceptibility and magnetization data presented in the main text of the manuscript, obtained using the methods described in Experimental section. In particular, we present a comparison of the original magnetic susceptibility data together with the related magnetic background measurements. All magnetic background data shown in Figs. \ref{fig:Zn_BIF_MvT_bgd}-\ref{fig:Ni_BIF_AC} are the direct measurement of the straw and plastic wrap done before the measurements of magnetic properties of a sample. The measurement parameters of the background and the sample were kept the same with the exception of a lower point density for the background. We also present here magnetic susceptibility measurements for Cu- and Co-BIF extended down to 0.4~K using the $^3$He option in QD MPMS. For data measured in the 0.4-2 K range, background magnetization was assumed to differ very little from the magnetization measured at T = 2 K in the standard $^4$He option of the MPMS and thus a constant background was subtracted from magnetization vs. temperature data. The constant used in subtraction was the background magnetization at T = 2 K, which is 3 orders of magnitude smaller than the sample magnetization of Cu- and Co-BIF at the same temperature. For magnetization vs. applied magnetic field at T = 0.4 K, the magnetization vs. field background measured at T = 2 K in the MPMS $^4$He option was subtracted from the data.

\subsection{Zn-BIF}

\begin{figure}[H]
    \includegraphics[width=\linewidth]{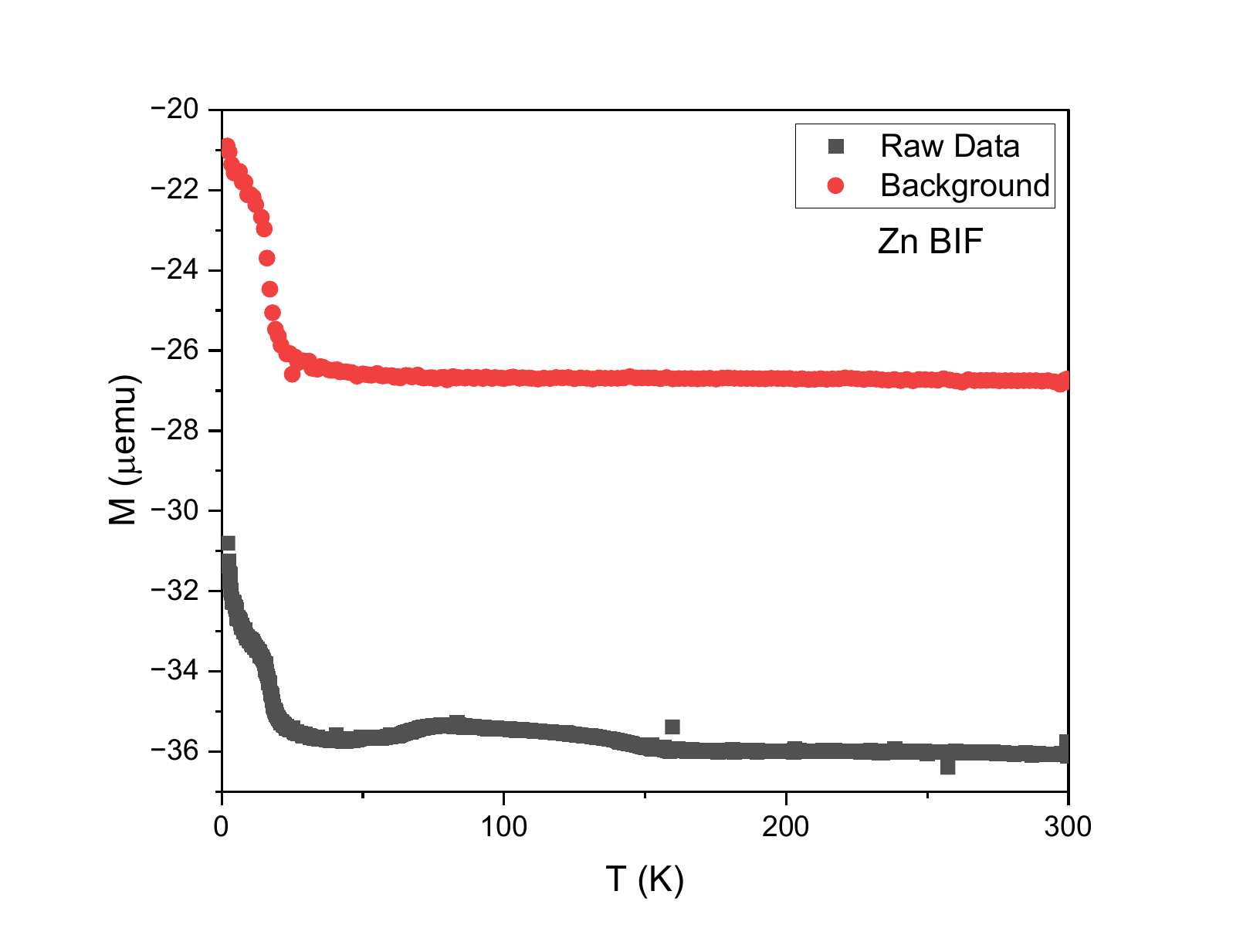}
    \caption{Raw magnetization  of Zn-BIF and plastic wrap/straw background vs. temperature, measured with 1000 Oe applied field.}
    \label{fig:Zn_BIF_MvT_bgd}
\end{figure}

 Fig. \ref{fig:Zn_BIF_MvT_bgd} shows raw magnetization data for 22.0 mg Zn-BIF and the plastic wrap/straw background data.  After subtraction, magnetization was scaled to molar susceptibility using a molar mass of 1354580 mg and an applied field of 1000 Oe, these data are presented in Fig. 2 of the main manuscript.

\newpage

\subsection{Cu-BIF}

\begin{center}
\begin{figure}[h]
    \includegraphics[width=1.45\linewidth]{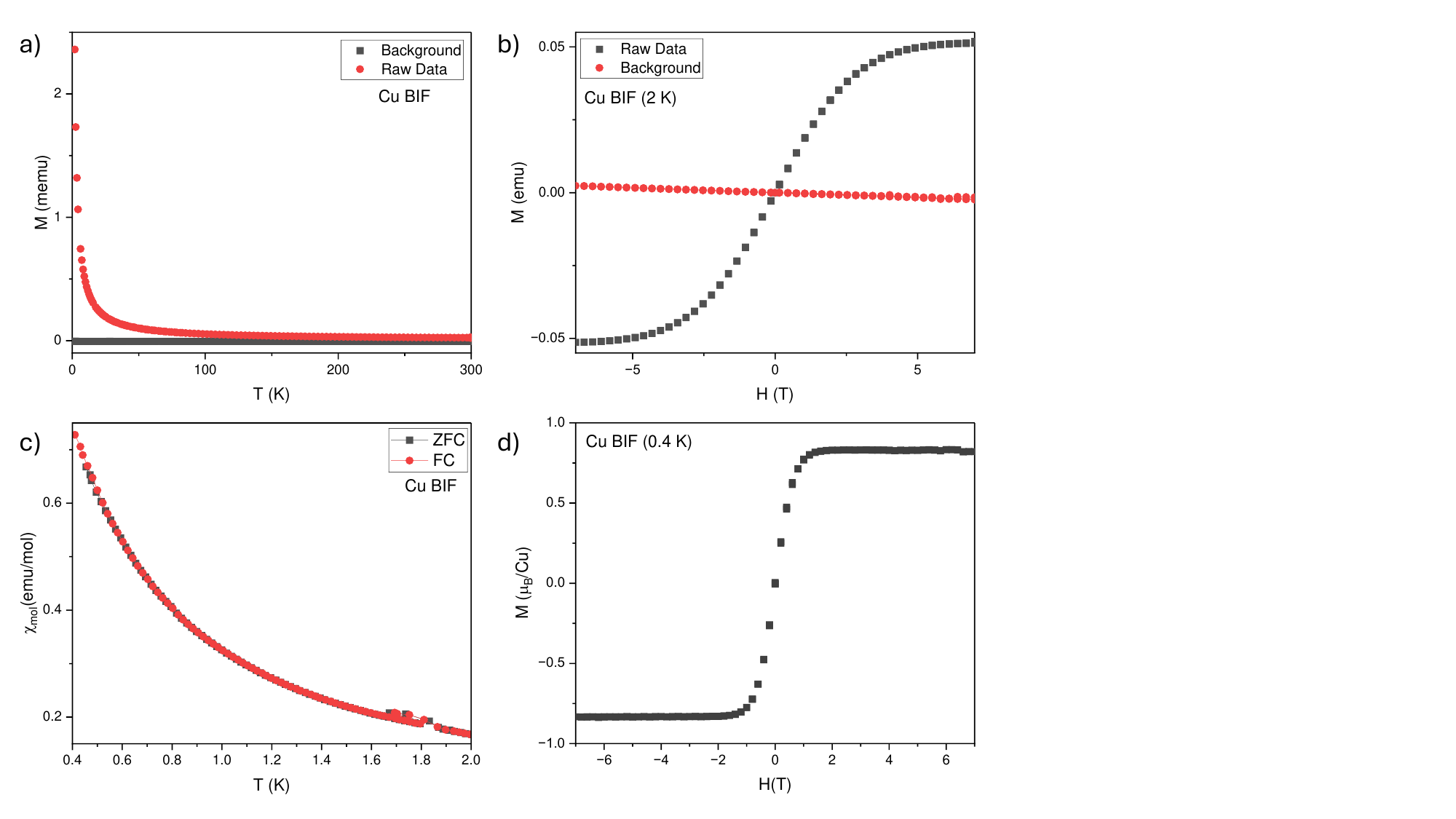}
    \caption{Magnetization of Cu-BIF and plastic wrap/straw background vs. temperature, measured with 1000 Oe applied field (a). Magnetization of Cu-BIF and plastic wrap/straw background vs. applied magnetic field, measured at T = 2 K (b). Zero-field cooled and field cooled magnetization of Cu-BIF measured from 0.4 K to 2 K with the MPMS $^3$He insert (c). Magnetization of Cu-BIF and plastic wrap/straw background vs. applied magnetic field, measured at T = 0.4 K with the MPMS $^3$He insert (d).}
    \label{fig:Cu_BIF_figs}
\end{figure}
\end{center}

Background and raw magnetization data are shown in Fig. \ref{fig:Cu_BIF_figs} for (a)  magnetization vs. temperature measurements for Cu-BIF sample with mass of  5.96 mg  and (b) magnetization vs. applied magnetic field measurements Cu-BIF sample with mass of 3.77 mg. After subtraction, magnetization was scaled to molar susceptiblity using a molar mass of 2077420 mg and an applied field of 1000 Oe. The scaled data are shown in Fig. 3 of the main manuscript. For magnetization vs. temperature measurements of the 3.77 mg sample in the 0.4-2K range, shown in panel c), a constant background of $-4.12*10^{-6}$ emu, the magnetization of the straw and plastic wrap background at T = 2 K, was subtracted before scaling. For magnetization at T = 0.4 K shown in panel d), the background magnetization at T = 2 K shown in panel b) was subtracted before scaling. 

\newpage

\subsection{Co-BIF}

\begin{figure}[h]
    \includegraphics[width=1.45\linewidth]{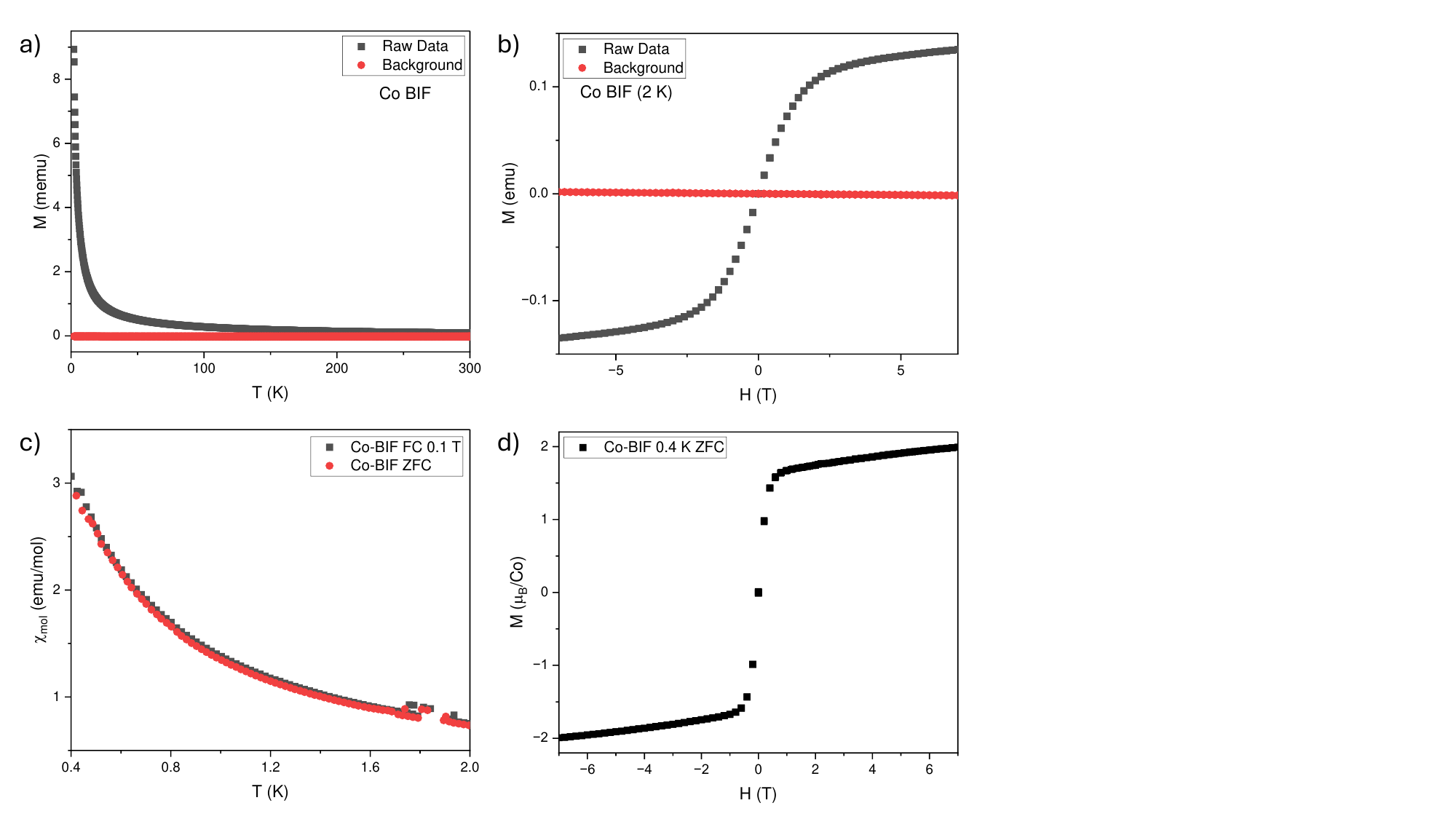}
    \caption{Magnetization of Co-BIF and plastic wrap/straw background vs. temperature, measured with 1000 Oe applied field (a). Magnetization of Co-BIF and plastic wrap/straw background vs. applied magnetic field, measured at T = 2 K (b). Zero-field cooled and field cooled magnetization of Co-BIF measured from 0.4 K to 2 K with the MPMS He3 insert (c). Magnetization of Co-BIF and plastic wrap/straw background vs. applied magnetic field, measured at T = 0.4 K with the MPMS He3 insert (d).}
    \label{fig:Co_BIF_figs}
\end{figure}

Background and raw magnetization data are shown in Fig. \ref{fig:Co_BIF_figs} for Co-BIF with mass of 5.57 mg . After subtraction, magnetization was scaled to molar susceptiblity using a molar mass of 484990 mg and an applied field of 1000 Oe. The scaled data are shown in Fig. 4 of the main manuscript. For magnetization vs. temperature measurements in the 0.4-2K range, shown in panel c), a constant background of $-1.6*10^{-5}$ emu, the magnetization of the straw and plastic wrap background at T = 2 K, was subtracted before scaling. For magnetization at T = 0.4 K shown in panel d), the background magnetization at T = 2 K shown in panel b) was subtracted before scaling. 

\newpage

\subsection{Ni-BIF}

\begin{figure}[hb!]
    \includegraphics[width=1.45\linewidth]{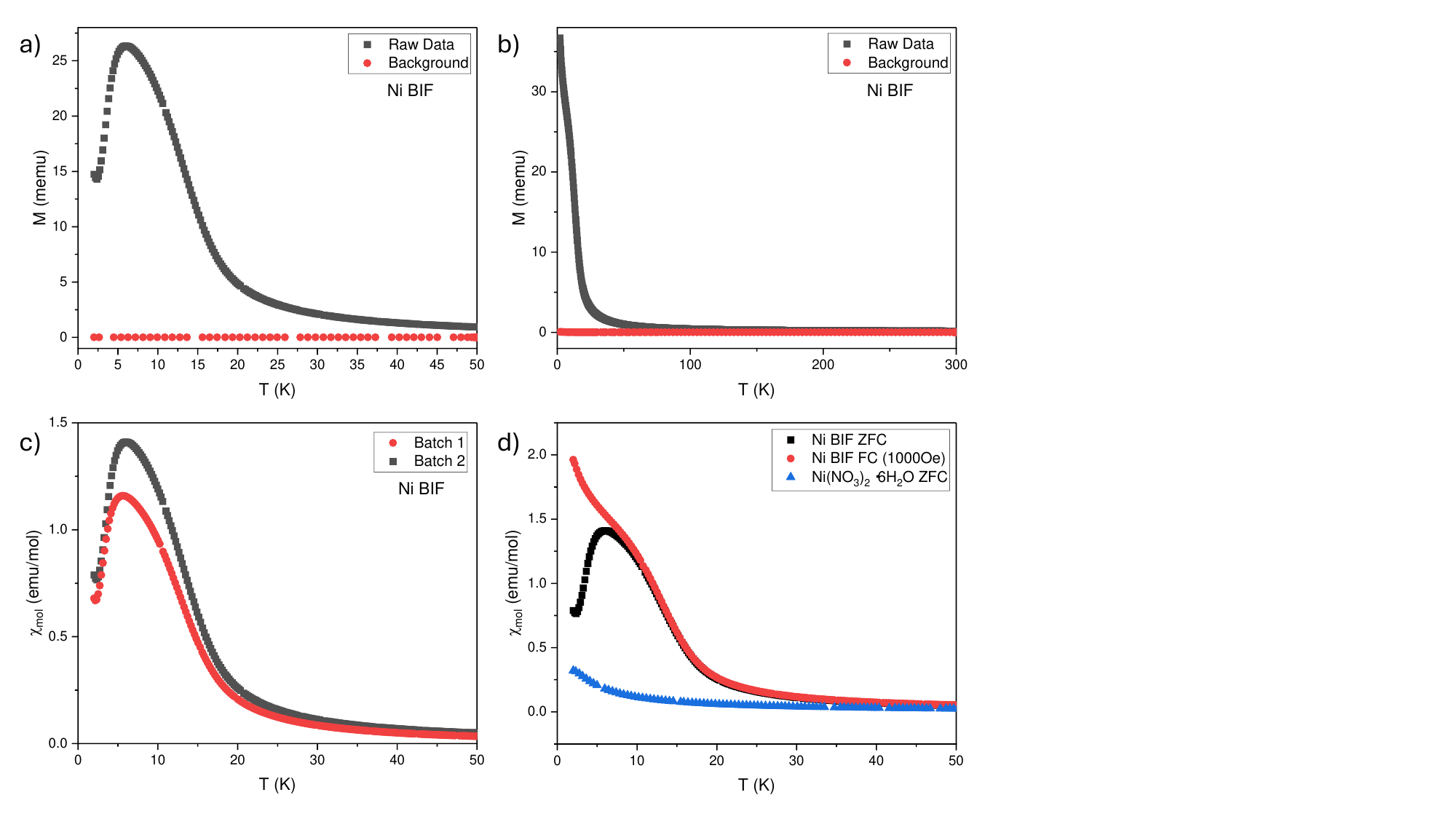}
    \caption{Zero-field cooled magnetization of Ni-BIF and plastic wrap/straw background vs. temperature, measured with 1000 Oe applied field (a). Field cooled magnetization of Ni-BIF and plastic wrap/straw background vs. temperature, measured with 1000 Oe applied field (b). Zero-field cooled magnetic susceptibility vs. temperature of two different batches of Ni-BIF, showing reproducibility. Peak temperatures differ by less than 0.5 K (c). Zero-field cooled and field cooled susceptibility of Ni-BIF, and zero-field cooled susceptibility of synthesis precursor Ni(NO$_3$)$_2 \cdot$ 6H$_2$O, showing simple paramagnetic behavior (d).}
    \label{fig:Ni_BIF_MvT}
\end{figure}

Two samples of Ni-BIF from different batches were measured, sample 1 (batch 1) with mass of 1.67 mg and sample 2 (batch 2) with mass of 9.05 mg. Background and raw magnetization data for sample 2 are presented  in Fig. \ref{fig:Ni_BIF_MvT}. After subtraction, magnetization was scaled to molar susceptibility using a molar mass of 484750 mg and an applied field of 1000 Oe. The scaled data are shown in Fig. 5 of the main manuscript. Panel c) shows a comparison of magnetic susceptibility measurements for these two samples. The difference in magnitude between the two samples may be attributed to an error in mass determination. Furthermore, the disordered nature of the interpretation discussed in the main text naturally suggests the possibility of small differences between batches. However, the results agree on the peak in susceptibility at around 6 K within 0.5 K. Panel d) compares magnetic susceptibility of sample 2 with that of  the synthetic precursor Ni(NO$_3$)$_2 \cdot$ 6H$_2$O, which displays paramagnetic behavior down to T = 2 K. This demonstrates that the magnetic behavior observed in Ni-BIF is robust and intrinsic to the BIF structure. 

\begin{figure}
    \includegraphics[width=\linewidth]{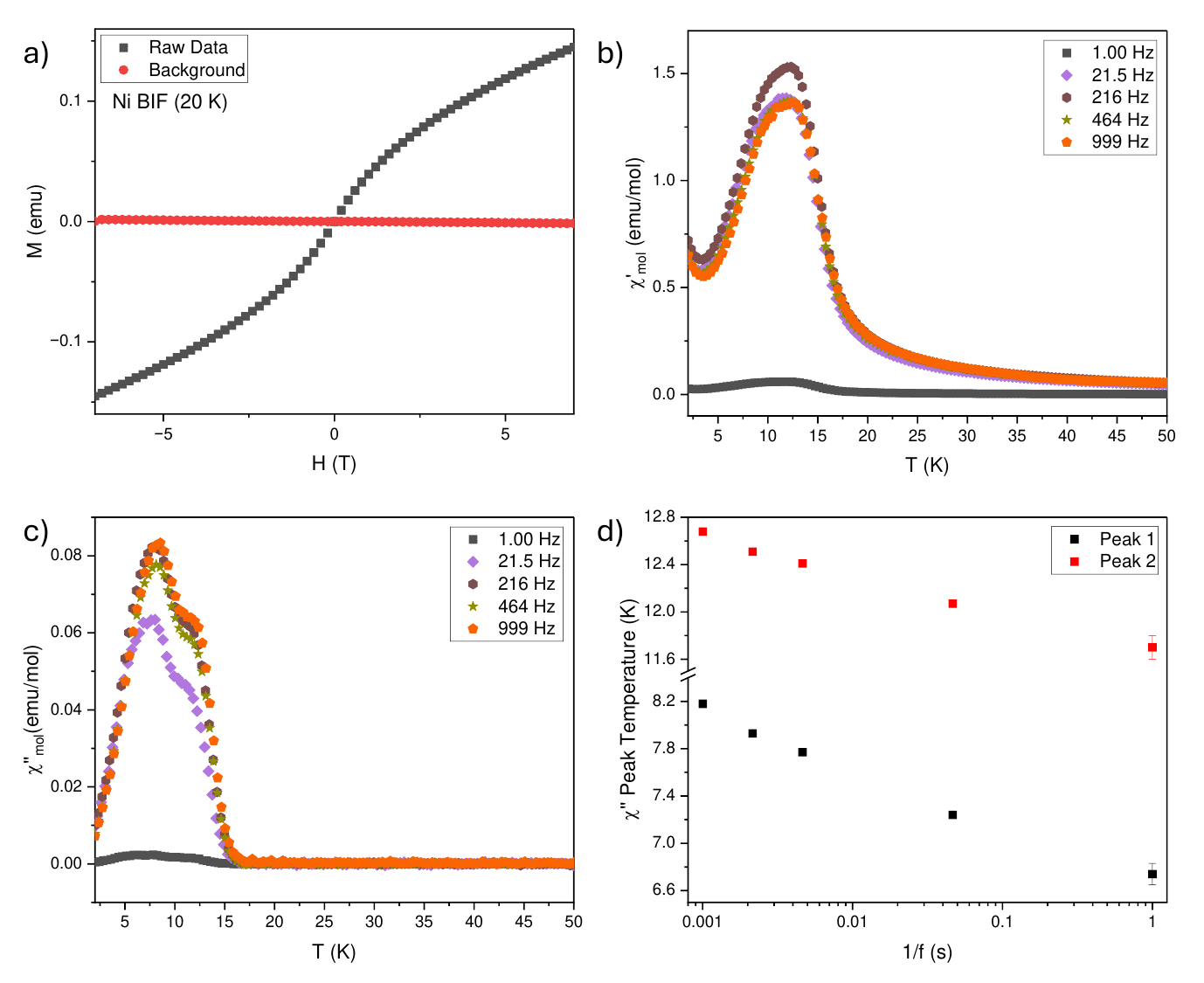}
    \caption{Magnetization of Ni-BIF and plastic wrap/straw background vs. applied magnetic field, measured at T = 20 K (a). Untreated in-phase component of molar AC susceptibility as shown in Fig. 8 of the main text (b). Out-of-phase component of AC susceptibility (c). Peak temperatures of two Gaussian peaks fit to out-of-phase AC susceptibility data (d).}
    \label{fig:Ni_BIF_AC}
\end{figure}

Fig. \ref{fig:Ni_BIF_AC} shows the raw sample and background magnetization measured for sample 2 of Ni-BIF at a temperature of 20 K from H = -7 to 7 T in panel a). Panels b) and c) show the raw in-phase and out-of-phase molar AC susceptibility of sample 2 of Ni-BIF measured with a 1 Oe applied AC H-field and no DC H-field at 5 different frequencies. No background was subtracted from the Ni-BIF AC susceptibility data, given that the DC susceptibility of the sample is at minimum 2 orders of magnitude larger than that of the background in this temperature range and that the features of interest are the positions of peaks in the out-of-phase signal, which will not be affected by a small diamagnetic background. {The in-phase susceptibility data shown in Fig. 8 of the main text is normalized to the 30-50 K $\chi'$ data at 999 Hz for clarity. The multiplicative scaling factors are: 28.17 (1 Hz),  1.14 (21.5 Hz), 0.99 (216 Hz), 1.04 (464 Hz), 1.00 (999 Hz).} The out-of-phase susceptibility peaks display {frequency-dependent peak temperatures and can be} fit by a sum of two Gaussian peaks using the nonlinear least-squares scipy.curve\_fit function in Python. Peak temperatures as a function of inverse frequency (measurement time) are displayed in panel d) and show a clear decrease upon reduction of the AC frequency. Error bars are obtained from the square root of the diagonal elements of the resulting covariance matrix.



